\documentclass[conference]{IEEEtran}
\makeatletter
\def\footnoterule{\relax
  \kern-5pt
  \hbox to \columnwidth{\hfill\vrule width 0.8\columnwidth height 0.4pt\hfill}
  \kern4.6pt}
\makeatother
\usepackage{cite}
\ifCLASSINFOpdf
   \usepackage[pdftex]{graphicx}
   \graphicspath{{../pdf/}{../jpeg/}}
   \DeclareGraphicsExtensions{.pdf,.jpeg,.png}
\else
   \usepackage[dvips]{graphicx}
   \graphicspath{{../eps/}}
   \DeclareGraphicsExtensions{.eps}
\fi
\usepackage{siunitx}
\usepackage{multirow}
\usepackage{xcolor}
\definecolor{redcolor}{rgb}{1, 0, 0}
\usepackage{subfiles}
\usepackage{makecell}
\usepackage{color,soul}
\usepackage{amsmath}
\usepackage{multirow}
\usepackage[normalem]{ulem}
\IEEEoverridecommandlockouts

\begin{document}
\setlength{\textfloatsep}{5pt}

\title{Analog Seizure Detection for Implanted Responsive Neurostimulation}

\author{
    \IEEEauthorblockN{
    Abbas A. Zaki\IEEEauthorrefmark{1}, 
    Noah C. Parker\IEEEauthorrefmark{1},
    Tae-Yoon Kim\IEEEauthorrefmark{1},
    Sam Ishak\IEEEauthorrefmark{1}, 
    Ty E. Stovall\IEEEauthorrefmark{1},
    Genchang Peng\IEEEauthorrefmark{1}, \\
    Hina Dave\IEEEauthorrefmark{2},
    Jay Harvey\IEEEauthorrefmark{2},
    Mehrdad Nourani\IEEEauthorrefmark{1},
    Xuan Hu\IEEEauthorrefmark{1},
    Alexander J. Edwards\IEEEauthorrefmark{1}, 
    Joseph S. Friedman\IEEEauthorrefmark{1}}
    
    \IEEEauthorblockA{\IEEEauthorrefmark{1}Department of Electrical and Computer Engineering, The University of Texas at Dallas, Richardson, TX
    \\\{abbas.zaki, noah.parker1, taeyoon.kim, sam.ishak, ty.stovall, gxp170004, \\ nourani, xuan.hu, alexander.edwards, joseph.friedman\}@utdallas.edu}
    \IEEEauthorblockA{\IEEEauthorrefmark{2}Department of Neurology, The University of Texas Southwestern Medical Center, Dallas, TX
    \\\{hina.dave, jay.harvey\}@utsouthwestern.edu}
}

\maketitle


\begin{abstract}
Epilepsy can be treated with medication, however, 30\% of epileptic patients are still drug resistive.  Devices like responsive neurostimluation systems are implanted in select patients who may not be amenable to surgical resection. However, state-of-the-art devices suffer from low accuracy and high sensitivity. We propose a novel patient-specific seizure detection system based on na\"ive Bayesian inference using M\"uller C-elements.  The system improves upon the current leading neurostimulation device, NeuroPace's RNS\textsuperscript{\tiny\textregistered} by implementing analog signal processing for feature extraction, minimizing the power consumption compared to the digital counterpart. 

Preliminary simulations were performed in MATLAB, demonstrating that through integrating multiple channels and features, up to $98\%$ detection accuracy for individual patients can be achieved.
Similarly, power calculations were performed, demonstrating that the system uses $6.5 \mu W$ per channel, which when compared to the state-of-the-art NeuroPace system would increase battery life by up to $50 \%$.

\end{abstract}

\renewcommand\IEEEkeywordsname{Keywords}

\section{Introduction}
As one of the most common neurological conditions, epilepsy affects 50 million people worldwide \cite{WHO}. Epilepsy is characterized by recurrent seizures resulting from excessive electrical discharges in either focal or generalized regions of the brain \cite{WHO}.  Seizure types can vary from brief lapses of attention to muscle jerks and debilitating convulsions \cite{WHO}. Approximately 30\% of epileptic patients suffer from refractory or drug-resistant epilepsy \cite{drug_resistant_epilepsy_2018}. Surgical resection is the conventional approach to treating drug-resistant focal epilepsy. However, resective surgery is not an option for all patients suffering from drug-resistant epilepsy especially if the seizures lack a well-defined epileptogenic region or if the region is in the eloquent cortex/deep brain structures. Furthermore, some patients continue to have seizures even after surgical resection \cite{Assi2017}. 

As surgical resection is not always possible or helpful, implantable neuromodulatoy devices delivering electrical stimulations to targeted regions of the brain are the next best choice to reduce or mitigate seizures \cite{Neuromodulation_treatment_2020,invasive_neurostimulation_epilepsy_2020,drug_resistant_epilepsy_2018}.  Two intracranial options include deep brain stimulation (DBS) and responsive neurostimulation.  The currently FDA-approved DBS device is an open-loop system acting similar to a pace-maker offering regular pulses with no feedback from sensors in the brain.  In contrast, responsive neurostimulation devices measure neural activity and offer stimulation when seizure activity is detected.  Historically there has been a two-fold trade off between detecting as many seizures as possible and not overstimulating the patient without compromising battery life.  In the extreme cases, embedded deep leaning models for seizure detection have been proposed which would be very power hungry, and, on the other extreme, some systems have limited to no seizure detection relying on over-stimulation to catch more seizure activity.

We propose here a solution to this problem taking advantage of the fact that seizure activity is often distributed in various networks throughout the brain.  We propose a solution using multiple low-power inferences from several EEG channels, combining potentially disagreeing evidences with an extremely low-power Bayesian inference engine comprising C-elements and stochastic bit-streams \cite{Friedman_C-Element_BI-2016}.  Preliminary simulations are performed demonstrating high detection accuracy, and a high-level power calculation is performed demonstrating that our detection engine can increase battery life by up to 50\% as compared with the state-of-the-art RNS\textsuperscript{\tiny\textregistered} device developed by NeuroPace. 

\section{Background}

\subsection{Previous Seizure Detection Solutions}

Seizure detection and treatment have been of great interest in recent years.  Implantable devices are of special interest in the clinical world because they can greatly improve the quality of life for patients dealing with chronic clinical seizures.  Most notably, NeuroPace's RNS\textsuperscript{\tiny\textregistered} system stands out for being the only FDA-approved responsive neurostimulation system on the market.  NeuroPace's system comprises a battery, controller, and one to two probes for EEG data acquisition and neurostimulation, in which charge controllably passes through brain tissue between selected contacts on the probes. 

For being the first system to market in this field, Neuropace's RNS\textsuperscript{\tiny\textregistered} system is outstanding{\textemdash}but it has its limitations. Notably, its method for seizure detection is not stellar, evaluating individual low-precision features to detect seizures.  This is intentional to minimize power usage. However, one major limitation of this is that as the detection scheme is not highly accurate, clinicians end up tuning these systems over months to years to be over-sensitive and to overstimulate the brain.  NeuroPace claims that during clinical studies, the device would stimulate patients roughly 600 times per day for less seizure-active patients and up to 2000 times a day for patients with increased seizure activity.  Though there are no known clinical drawbacks to over-stimulation, it is wasteful from a power management perspective, and based on a rough calculation from numbers in \cite{NeuroPace} stimulation uses up to 20\% of the battery life depending on the patient's neural activity and the stimulation parameters.  With a better inference engine, hyper-sensitivity could be reduced and battery life could be significantly extended while retaining the demonstrated  benefits of responsive neurostimulation systems. 

For maximizing the battery life of an implantable seizure treatment device, then, the difficulty lies in balancing the energy usage from high-accuracy feature evaluation and the energy lost to over-stimulation.  There are a number of resources demonstrating accuracy and energy for different types of features, and the results are pretty uniform: looking at single channel, exotic features (often implemented with Digital Signal Processing (DSP) blocks or neural networks) can achieve stellar accuracies but are very power hungry \cite{AFullyInteg8ChCL-Chen-2014,APersonalEAIEEG-Pinto-2021}.  Simultaneously, low-power features can have significantly lower accuracy \cite{An8ChannelSEEGASWPS-Yoo-2013,A16ChannelPSSOaT-Altaf-2015}, though when multiple low-power features are analyzed jointly, that accuracy can be improved \cite{An8ChannelSEEGASWPS-Yoo-2013,A16ChannelPSSOaT-Altaf-2015}.  Thus, a low power solution can be found in integrating evidences from multiple channels evaluated with low-power features. The question remaining is whether there is a low-power solution to integrate these potentially conflicting, low-power evidences from multiple channels simultaneously.  There is.


\subsection{Bayesian Inference with M\"uller C Elements}
\label{sec:C-E}

\begin{figure}[t]
    \centering
    \includegraphics[width=250pt]{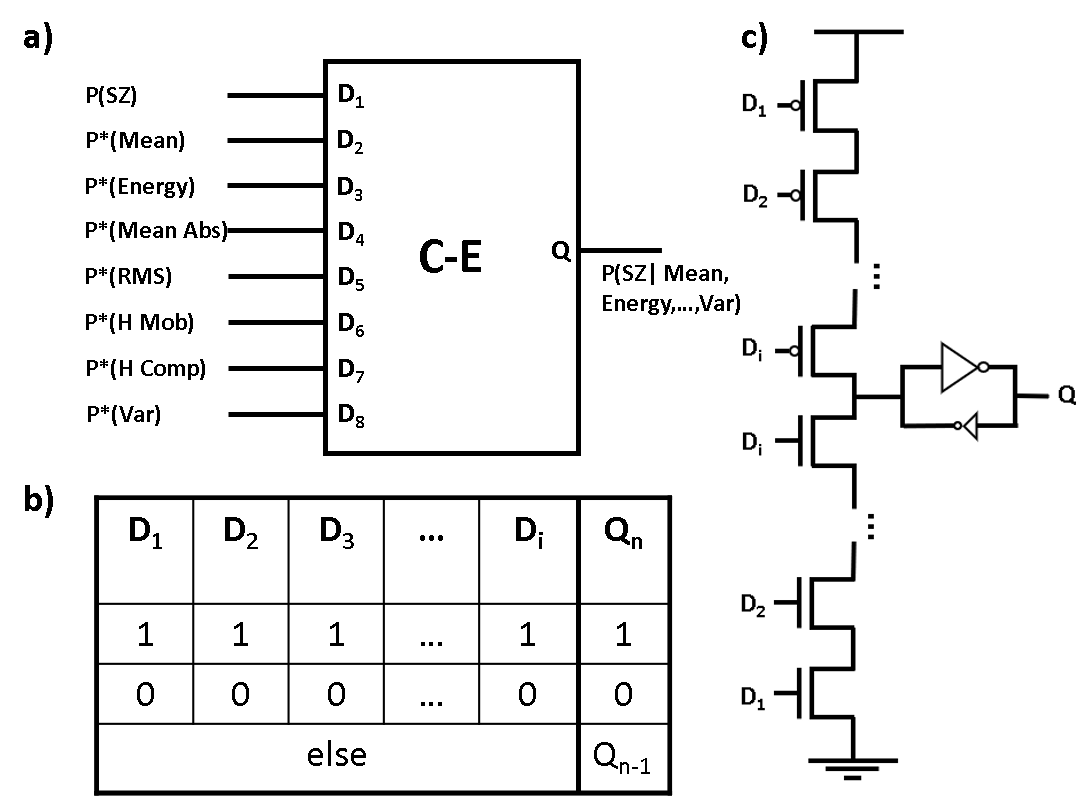}
    \caption{\textbf{a)} Multi-input C-Element for Bayesian inference. \textbf{b)} C-Element truth table. \textbf{c)} C-Element circuit implementation.}
    \label{fig:C-Elements-features}
\end{figure}

We previously proposed an extremely low-power implementation of Bayesian inference using C-Elements (C-Es) and stochastic bitstreams \cite{Friedman_C-Element_BI-2016}. A diagramatic representation of the M\"uller C-element and its Truth table can be seen in Fig.  \ref{fig:C-Elements-features}.  In brief, Bayesian inference computes the probability of an event $V$ given the $i$ observed evidences $E^m_1, E^m_2, ... E^m_i$ where $m$ is the observed value of the corresponding evidence, so $P(E^{0.5}_3)$ corresponds to the probability that evidence 3 has the value of $0.5$.  With this notation, na\"ive Bayesian inference is formulated as:

\begin{multline}
P(V \mid {E^m_1,...,E^m_i}) \\
= \frac{P(V)\prod^i_{x=1} P^*(E^m_x)}{P(V)\prod^i_{x=1} P^*(E^m_x)+(P(\bar V))\prod^i_{x=1}(1-P^*(E^m_x))} ,
    \label{eq:BI}
\end{multline}

where $P^*(E^m_x)$ is defined as

\begin{equation}
P^*(E^m_x)=\frac{P(E^m_x \mid V)}{P(E^m_x \mid V) + P(E^m_x \mid \bar V)} .
    \label{eq:PStar}
\end{equation}

Stochastic computing is a recently proposed methodology for approximate computing \cite{stochastic-computing}.  Real numbers in $[0,1]$ are encoded as the probability that at any given time the state of a stochastic bitstream will be $1$.  Stochastic computing offers new ways of computing mathematical operations; for instance, an AND gate can perform multiplication and a MUX can be used for weighted addition \cite{stochastic-computing}.  Most notably \cite{Friedman_C-Element_BI-2016} recognized that multi-inut a M\"uller C-E block in the stochastic regime can compute the function:

$$Q = \frac{\prod^i_{x=1} D_x}{\prod^i_{x=1} D_x + \prod^i_{x=1}(1-D_x)}$$

Which when the $D_x$ values are interpreted as $P(V)$ and $P^*(E^m_x)$, directly implements eq. (\ref{eq:BI}). Thus, in the stochastic regime a single multi-input C-E \cite{HOE20194, Friedman_C-Element_BI-2016} will directly implement Bayesian inference. In \cite{Friedman_C-Element_BI-2016} analysis was performed demonstrating that with minor accuracy loss, energy usage can be heavily reduced to the order of $nJ$ per-inference. 

Ultimately then, from a high-level view of the field, for high-accuracy and low-power seizure detection, multiple low-power features across different channels should be analyzed and integrated using low-power inference. We propose such a system which does not require high-power analog-to-digital conversion (ADC) and digital signal processing (DSP) blocks, by instead making use of alternative analog and stochastic blocks for energy-efficient, high-accuracy seizure detection.

\begin{figure*}[t]
    \centering
    \includegraphics[width=500pt]{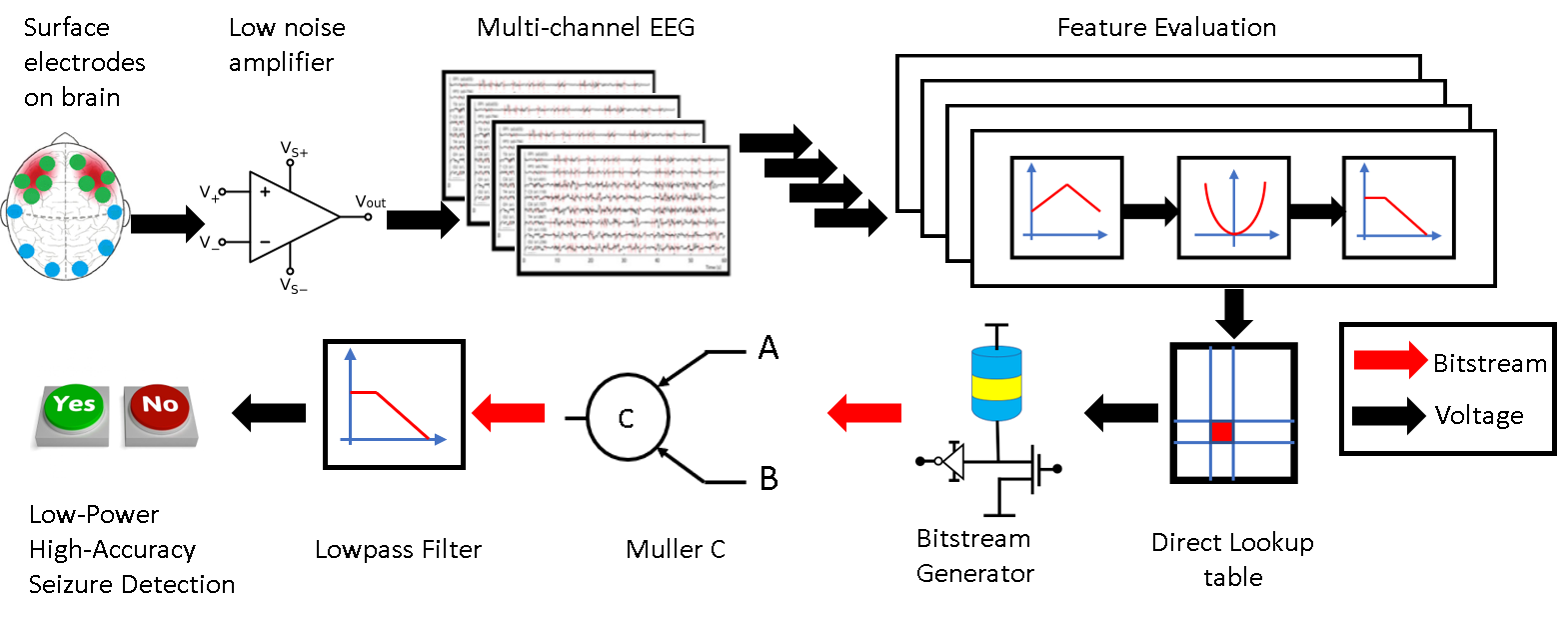}
    \caption{Data path from EEG signal acquisition to seizure detection.}
    \label{fig:dataflow}
\end{figure*}

\section{System Overview}
\label{sec:sys}

To enable an extremely low-power solution for seizure detection avoiding DSP computations and ADCs, analog processing blocks are used as much as possible.  Fig.  \ref{fig:dataflow} depicts the complete data path from EEG signal extraction to seizure identification.  

To minimize power usage, the system will not store or transmit EEG data; rather it's sole purpose will be to detect and treat seizures.  This heavily limits the adjustability of the system after implantation.  In this work, it is assumed that before surgery, copious information about the patient's seizure networks has been gathered, and proper EEG channels and associated features can be identified and tuned before implantation surgery.  From our analysis in section \ref{sec:inference_accuracy} the number of $(feature, channel)$ pairs that need to be extracted for high-accuracy detection varies from one to four depending on the patient. 

This pre-surgery study of the patient will govern the placement of probes in the brain.  EEG voltage signals range from hundreds of micro-volts to tens of milli-volts and information is encoded in frequency bands of tens of hertz to hundreds of hertz.  These somewhat low-frequency and low-voltage signals can be amplified by a low-noise differential amplifier to produce an analyzable rail-to-rail signal.  One amplifier will be needed per channel analyzed. 

Once these signals are obtained, they may be fed into one of several types of features including \textit{mean}, \textit{energy}, \textit{mean of absolute value, line lenght, Hjorth mobility, etc.} each of which has a low-power approximate analog implementation to avoid the high power cost of DSP.  Many of these features require integrating information over a window of time.  In our initial discrete simulations, it was found that a window of roughly five seconds contained enough information to identify a seizure.  Thus, for the analog implementations of these features, time constants for decaying integration were chosen on the order of five seconds.

Emerging from each feature will be a level voltage $v_f(t)$ corresponding to the feature value at time $t$.  This voltage must be prepared for Bayesian inference by being transformed to represent $P^*(E^v_f)(t)$ introduced in eq. \ref{eq:PStar}.  During the pre-surgery study period, the function $p:v_f\rightarrow P^*(E^v_f)$ will be defined for each useful feature for each channel indicating how likely a seizure is, given the observed feature value $v_f(t)$.  The function $p$ will be hardcoded into a voltage-in, voltage-out \textit{lookup table} (LUT) which can be implemented through multiple means.  Ideally, it can be closely matched with a low-power analog implementation.  Likewise, we demonstrate below in Sec. \ref{sec:ideal_results} that only a small number of bins per feature need to be created for high-accuracy detection, so a series of low-power level detectors can be substituted without loss of accuracy.  As stated above, the system will not be adjustable post-implantation, so an analog lookup table may be possible here where it is not in other applications.

Once the voltages representing $P^*(E^v_f)(t)$ are obtained, they are directly fed into a voltage-controlled tunable random number generator (RNG) which produces stochastic bitstreams with a time-varying probability equal to $P^*(E^v_f)(t)$.  Multiple low-footprint tunable RNGs have been recently shown some utilizing super-paramagnetic magnetic tunnel junctions (MTJs) to provide entropy \cite{Zhou2008, Yang2015, Mathew2016, Vodenicarevic2017, P-Bit_generators-RFaria-2017, Camsari2017-Stochastic-P-Bits-MTJ}.

These bitstreams are fed asynchronously into the C-E(s) to perform na\"ive Bayesian inference as described in \cite{Friedman_C-Element_BI-2016}. The output of this network will be a final stochastic bitstream representing the probability that the patient is currently in an ictal (seizure) state.  Passing this bitstream through a passive low-pass filter is sufficient to convert this bitstream to a level voltage.  A voltage above a pre-determined threshold will represent a seizure state.  After the threshold is applied, this Boolean signal may be fed into a treatment subsystem (not shown) to stimulate the seizure network and stifle the seizure before it causes adverse symptoms.

All of the data-path blocks described above are designed to be as energy efficient as possible while still providing high-accuracy detection.  The following two sections demonstrate preliminary results showing high inference accuracy and very low power respectively.

\begin{figure}[b]
    \centering
    \includegraphics[width=250pt]{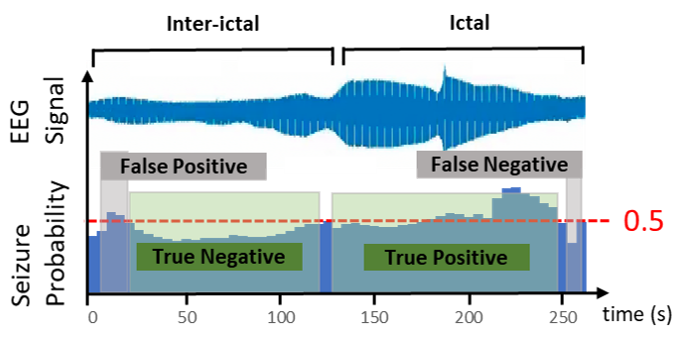}
    \caption{Seizure Detection with Discrete Windows. For each 5-second window, the system identifies whether there is a seizure or not for both Ictal and Inter-Ictal states keeping false negatives and false positives to a minimum.}
    \label{fig:Seizure Window}
\end{figure}

\begin{figure}[t]
\centering
    \includegraphics[width=250pt]{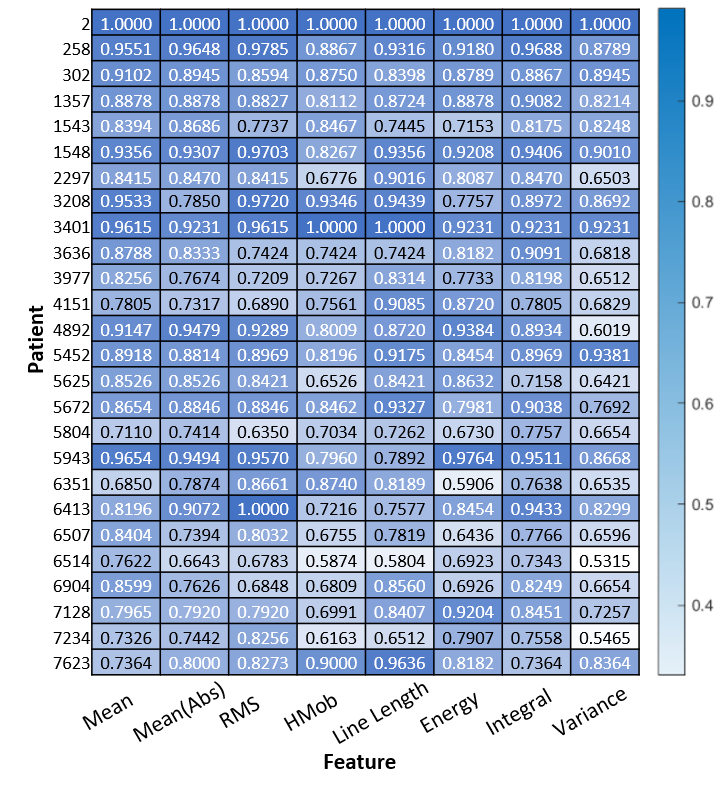}
    \caption{Windowed Single Feature Performance. The highest detection accuracies for each feature for each patient spanning the 24 EEG channels are depicted here, demonstrating that simple features can be used to obtain high detection accuracies for some patients.}
    \label{fig:ideal-table}
\end{figure}

\section{Inference Accuracy}
\label{sec:inference_accuracy}

While intracranial EEG data would be ideal for an implantable seizure detection system, the lack of publicly available databases necessitates the use of scalp EEG data to demonstrate the system's viability as a low-power detection device. It is important to note though that scalp EEG signals contain more noise, as the electric potentials being measured face interference from bone and skin; therefore, a system that can reliably detect seizures with low power on scalp EEG data would be expected to perform equally well, if not better, on intracranial EEG data albeit with some modifications to thresholds.
As such, the Temple University Seizure Corpus (TUSZ) \cite{TUH-Database} was chosen to evaluate the potential effectiveness of the proposed system. The corpus provides one of most expansive collections of seizure and non-seizure EEG recordings and contains clinical annotations marking periods of EEG data that contain seizures, granting a nice interface  to demonstrate the high detection accuracies we were able to achieve with our methodology.

\subsection{Ideal Windowing and Features}
\label{sec:ideal_results}

Initial simulations were performed in the digital domain.  Patients from the TUSZ database were selected for study if they had seizures that lasted 60 seconds or longer and had non-seizure activity that lasted 60 seconds.  These patients were evaluated individually to determine the effectiveness of our proposed system. Each EEG signal was processed using a bipolar Temporal Central Parasagital (TCP) linked-ears reference montage \cite{TUH-Database} so that each EEG recording contained 22 channels of EEG activity. 

The seizure and non-seizure activity for each patient was then grouped into 5-second windows and labeled as containing seizure or non-seizure data as shown in Fig. 3.
For each patient, each feature was evaluated on each channel for each 5-second window.
None of the features rely on Fourier or wavelet transforms as used in other works to avoid incurring high computational costs associated with each transformation into the frequency/spectral domain. 

From here the calculated features were evaluated according to how well they could be used to detect seizures.  
For the model, a probability table corresponding to the probability distribution function for each feature-channel pair was generated from the training set.  This table gives the probability of observing a specific feature value in an ictal vs. inter-ictal state, or $P(feature | ictal)$. A non-uniform binning algorithm was used here to ensure that there was a statistically useful amount of data in each bin and to limit the overall number of bins.

\begin{figure}[b]
    \centering
    \includegraphics[width=250pt]{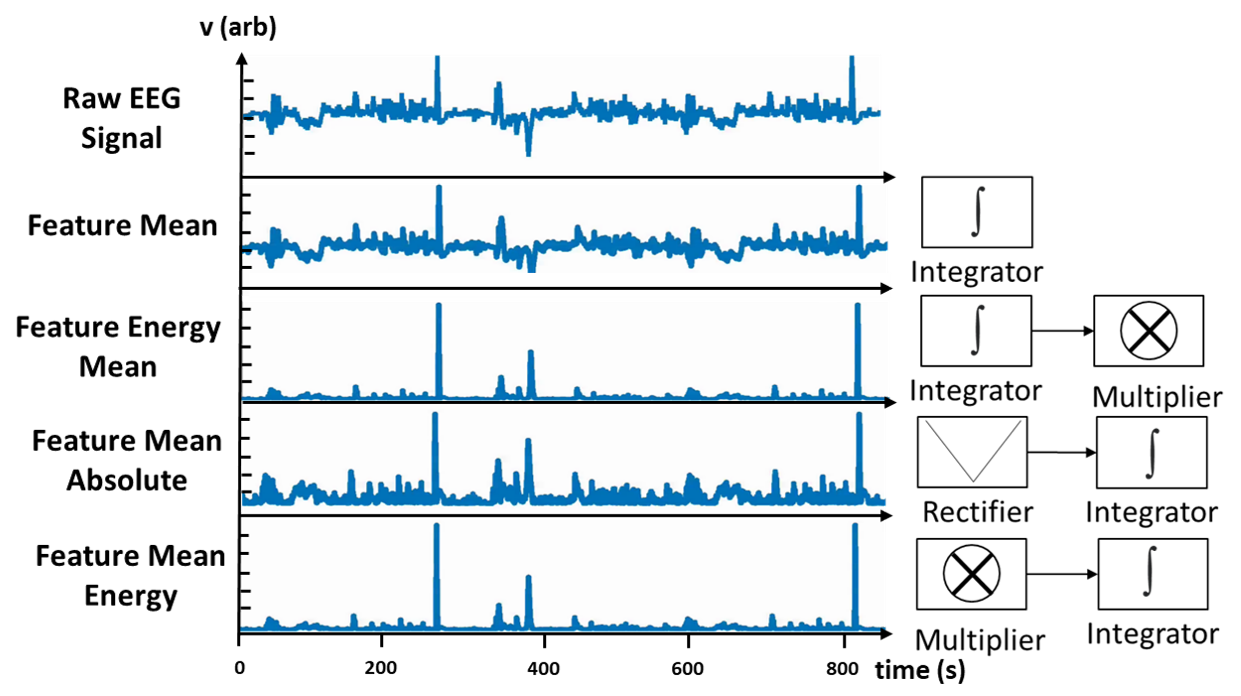}
    \caption{Analog features.  A raw EEG signal along with extracted features is shown.  Simulink blocks used to implement these features are likewise depicted.}
    \label{fig:analogblock}
\end{figure}

Na\"ive Bayesian inference was then computed directly from this probability distribution to classify each window in the testing set as showing either ictal or inter-ictal data.  The J-statistic was used to consolidate false-negatives, false-positives, etc. into a single metric on $[-1, 1]$ where a value of $-1$ implies no classifications were made correctly, and a value of $1$ corresponds to perfect classification.  Results for each patient are shown in Fig. \ref{fig:ideal-table}.  For each feature, only the J-statistic for the channel with the highest performance is shown for brevity.  High classification accuracies were obtained for some feature-channel combinations for some patients indicating the ability for these features to be used to classify focal seizures. 

\begin{figure}[t]
    \centering
    \includegraphics[width=250pt]{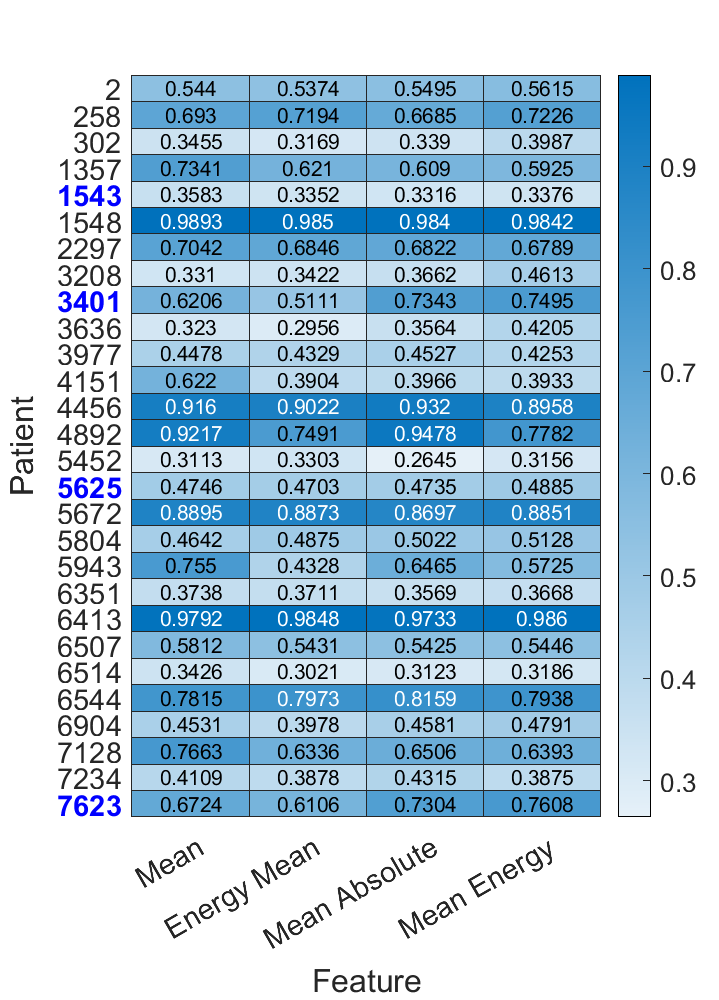}
    \caption{Analog Single Feature Performance. For each feature, only the highest accuracy obtained among the 24 EEG channels is depicted.  Patients 1543, 3401, 5625, and 7623 saw significant increases in accuracy when integrating information from multiple channels and features.}
    \label{fig:analog:heatmap}
\end{figure}

\begin{figure}[t]
    \centering
    \includegraphics[width=250pt]{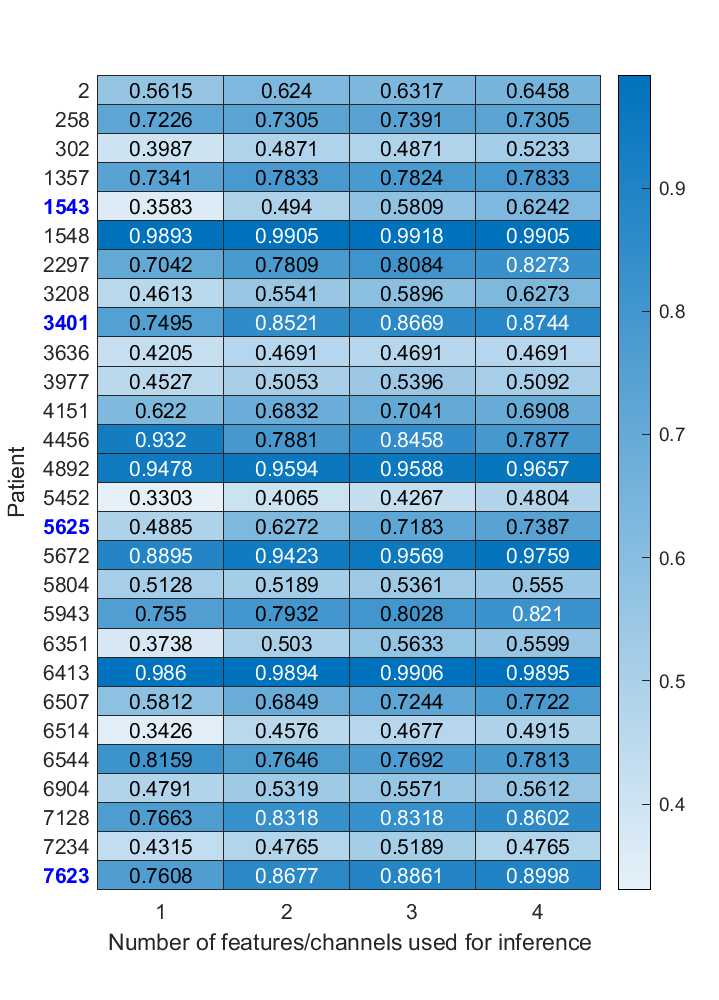}
    \caption{Analog Multi Feature Performance. Patients 1543, 3401, 5625, and 7623 saw significant increases in accuracy when integrating information from multiple channels and features.}
    \label{fig:analog:multifeature_heatmap}
\end{figure}

\begin{figure}[t]
    \centering
    \includegraphics[width=250pt]{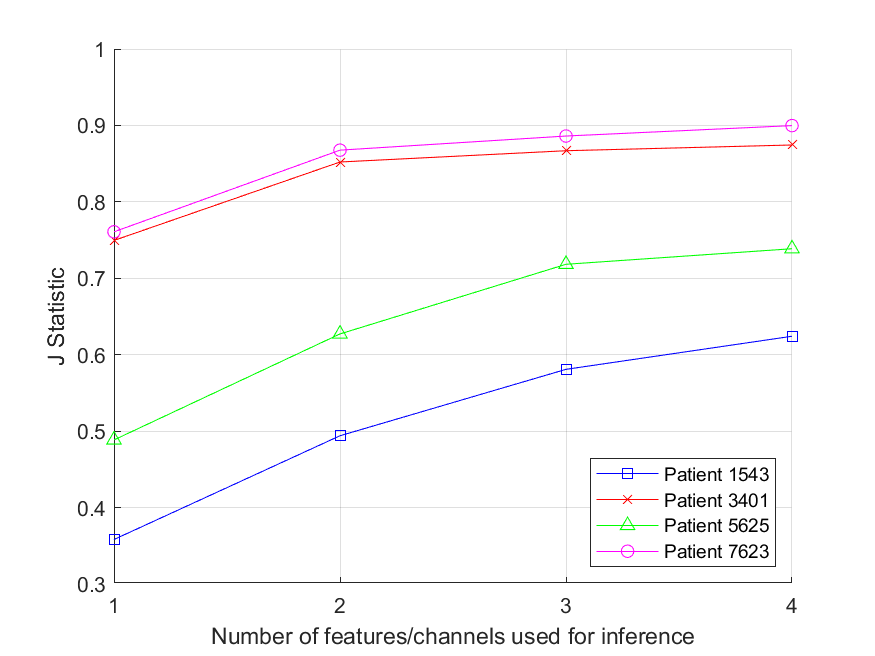}
    \caption{J Statistic vs Number of Features for selected patients. The accuracies for the patients highlighted in Figs. \ref{fig:analog:heatmap} and \ref{fig:analog:multifeature_heatmap} and  are tracked according to how many features were used in each evaluation.  Some patients saw up to a $30 \%$ increase in accuracy, and trends indicate that these accuracies will continue to increase with the further addition of features.}
    \label{fig:analog:highlighted_patients}
\end{figure}

\subsection{Analog Feature Simulation and Results}

To secure validate this result, the same simulations were performed in the analog domain using Matlab's Simulink where EEG signals were not discretely sampled into 5-second windows, but rather, the various time-constants of the feature circuits emulated a memory window of approximately 5-seconds.  Fig.  \ref{fig:analogblock} depicts how a number of these features transform the input EEG signals.
After circuit simulation was completed, the feature values were binned and used to perform na\"ive Bayesian inference as before.  For this experiment, a 5-fold cross-validation procedure was implemented as in \cite{Shoaran2018_EnEffClass_5FoldCV, TRUONG2018-ConvNN, Daoud2019-EESeizPredDeepLearn, Yang2020-SeizRev} wherein the data set was partitioned into five subsets.  Five trials were run on the data set, each time using a different subset of the data as the testing set and the other four subsets as the training set.  This was done to increase confidence in our results obtained from a relatively small data set for each patient.

Fig.  \ref{fig:analog:heatmap} displays these results showing agreement with the analysis in section \ref{sec:ideal_results}.  Note that some of the features from Sec. \ref{sec:ideal_results} were not simulated here, though all of these features have an approximate, low-power analog implementation.  It is expected that the results from the ideal versions of these features will extend to the analog domain as they do with the features presented here.

Classification based on a single $(feature, channel)$ pair may be effective for patients with very focal seizures, but is not effective for many patients whose seizures are distributed across the brain.  Analog domain simulations were extended to utilize multiple features across multiple channels simultaneously using na\"ive Bayesian inference.  All combinations of two features were evaluated to identify the highest performing pair.  While exhaustive evaluation of all possible three and four feature combinations was not feasible, a genetic algorithm optimizer was seeded with the highest performing two feature pair and used to find high performing three and four feature combinations. The results of these optimizer simulations is shown in Fig. \ref{fig:analog:multifeature_heatmap}.

Our results demonstrate the feasibility of high-accuracy seizure detection using low-power analog features and Bayesian inference.  $93\%$ of patients saw an increase in seizure classification accuracy using two feature-channel pairs, and a number of patients continued to see an increase in accuracy through the case of four feature-channel pairs, several of which are highlighted in Fig. \ref{fig:analog:highlighted_patients}. It is anticipated that combinations of five or more $(feature, channel)$ pairs can be used to increase these accuracies further.  The best performing patient saw a $30\%$ accuracy increase between one and four features.  

We have demonstrated then that for patients with seizure networks, multiple features and channels can and should be used to boost detection accuracy, decreasing false positives and ultimately the number of stimulations issued to the brain.  What remains to be shown here is that these high detection accuracies can be obtained in a sufficiently low-power context to ensure long battery life for the highest quality of life for the patient. 

\section{Power Analysis}

We obtain an upper bound on the power consumption of the detection system.  We perform power analysis of each individual block of Fig. \ref{fig:dataflow} with the system parameters in Table \ref{tab:system}. Assuming a battery with $3.3 Wh$ capacity as in \cite{NeuroPace}, the detection system is able to maintain high-accuracy detection using only $p \mu W$ of power.


\newcommand{\minus}{\scalebox{0.75}[1.0]{$-$}}
\renewcommand\theadalign{bc}
\renewcommand\theadfont{\bfseries}
\renewcommand\theadgape{\Gape[4pt]}
\renewcommand\cellgape{\Gape[4pt]}

\begin{table}[t]
\centering
\caption{System Parameters}
\resizebox{0.4\textwidth}{!}{%
\begin{tabular}{|c|c|}
\hline
\thead{Parameter}      & \thead{Value} \\ \hline
Supply Voltage         & 1.2 V   \\ \hline
Sampling Frequency     & 1000 Hz \\ \hline
Average stimulation current & 1 mA\\ \hline
Maximum stimulation current & 12 mA\\ \hline
Lead Resistance & 1200 ohms \\ \hline
Stimulation Pulse Width & 160 $\mu$s \\ \hline
Stimulation Period  & 5 ms \\ \hline
Average Stimulation Duration & 100 ms \\ \hline
Maximum Stimulation Duration & 5000 ms \\ \hline
Parasitic Capacitance  & 100 fF  \\ \hline
Load Capacitance       & 20 pF   \\ \hline
Battery Capacity       & 3.3 Wh   \\ \hline
\end{tabular}%
}
\label{tab:system}
\end{table}

\subsection{Signal Sensing and Pre-Amplification}

The first step of obtaining EEG signals is to detect brain activity through electrodes. After the electrodes detect the signals, a preliminary low-noise amplifier is needed to amplify $mV$ or $\mu V$ signals to sub-volt levels for the cascading circuits to efficiently and accurately process the EEG signals\cite{MullerAmplifier2012}. As in \cite{Xu2008,Zou2010,Denison2007}, the current consumption for the preliminary EEG signal amplification process process is less than $1\ \mu A$ per channel. As the EEG signals generally have an upper bandwidth limit of $100\ Hz$, sampling such EEG signals will not introduce notable extra energy with Nyquist sampling \cite{Zou2010}. The estimated power consumption for sensing and pre-amplification is therefore bound at $1.2\ \mu W$ per $(feature,channel)$.

\subsection{Feature Extraction}

To evaluate the power consumption of feature extraction circuits, we implement four feature evaluation methods as shown in Table \ref{tab:features} that are used for the simulations in Section \ref{sec:inference_accuracy}. As shown in the block diagram of Fig. \ref{fig:analogblock}, the core analog computing blocks for feature extraction are based on multipliers and op-amps. The simulations are performed in Cadence with 65 nm technology node with a 1.2 V supply voltage. Each simulation is run for 10 seconds for the power calculation with the input signal frequency up to $100\ Hz$, which matches EEG signal bandwidth. Table \ref{tab:features} summarizes the simulation results. As can be seen in Table \ref{tab:features}, the maximum power consumption of feature extraction per channel is $3.539\ \mu W$, and the average power consumption of feature extraction per channel is $2.523\ \mu W$. 



\begin{table}[t]
\centering
\caption{Power consumption of EEG feature extraction circuits}
\resizebox{0.35\textwidth}{!}{%
\begin{tabular}{|c|c|}
\hline
\thead{Feature name}                & \thead{Power}   \\ \hline
Feature\_mean                       & 1.995   $\mu$W  \\ \hline
Feature\_mean\_abs                  & 3.539   $\mu$W  \\ \hline
Feature\_mean\_energy               & 1.680   $\mu$W  \\ \hline
Feature\_energy\_mean               & 2.879   $\mu$W  \\ \hline
\end{tabular}%
}
\label{tab:features}
\end{table}

\subsection{Analog Lookup Table}
To convert the results from the feature extraction circuits to a voltage that modulates the switching probability of the RNGs, an analog-to-analog lookup table circuit is used.  As in Section \ref{sec:sys} this circuit will be different for each patient for each feature, so there will be various implementations depending on the function the circuit is representing.  In the worst case, a set of discrete level detectors will be used, and this is the circuit we analyze here in order to provide an upper bound for the power usage of the system.

A multilevel voltage reference circuit is used to generate both the referencing and driving voltages, while a selector network is used to arbitrarily map the input and output voltages. This circuit is simulated in GF 65nm technology; a eight-level analog lookup table circuit consumes $493.80 nW$ of power which scales linearly with the number of discrete input levels. The architecture of this circuit is preliminary, and it is anticipated that power consumption can be optimized further.  In our simulations in Section \ref{sec:inference_accuracy} approximately 40 discrete bins were used to group feature and probability data.  Using linear scaling then, we estimate a worst case power consumption of $2.469 \mu W$ per $(feature,channel)$ for this block.

\subsection{Stochastic Bitstream Generator}
\label{sec:TRNG}

As in \cite{Friedman_C-Element_BI-2016} and Section \ref{sec:C-E}, random number generators (RNGs) are required for the Bayesian inference circuit with C-elements. While conventional CMOS-based RNGs are also available for low-energy random bitstream generation \cite{Mathew2016,Yang2015,Zhou2008}, we propose to use stochastic p-bits \cite{Camsari2017-Stochastic-P-Bits-MTJ} which can be used to generate true random numbers with low energy cost \cite{Vodenicarevic2017} given the low energy barrier of the nanomagnets \cite{Camsari2017-Stochastic-P-Bits-MTJ}. As in \cite{Vodenicarevic2017}, the true random number generator (TRNG) consumes $20\ \text{fJ/bit}$. Assuming a 1000 Hz sample rate which will produce sufficiently long bitstreams for proper classification, power consumption per $(feature,channel)$ for the TRNG can be estimated as:
\begin{equation}
    P_{TRNG} =   E_{bit} \cdot f_s  = 20\ \text{pW},
\end{equation}
where $P_{TRNG}$ is the power consumption for the entire TRNG block of the proposed system, $E_{bit}$ is the energy required to generate one bit of random number, and $f_s$ is the sample rate.

Compared to other analog blocks within the proposed system, the RNG consumes significantly less power regardless of whether the RNGs are constructed by emerging technology devices or conventional CMOS circuits \cite{Mathew2016,Yang2015,Zhou2008}.

\subsection{M\"uller C-Elements}

The proposed system requires a multi-input C-element circuit as shown in Fig. \ref{fig:C-Elements-features}. As in \cite{Friedman_C-Element_BI-2016}, each two-input C-Element consumes $20 fJ$ per bit operation. Since the power consumed by the C-element blocks are more than three orders of magnitude less than the major power consumers of the system, the power consumption of one \textit{n}-input C-element for \textit{n} channels can be estimated to be the same as \textit{n} two-input for all channels, which can be normalized to one C-element per $(feature,channel)$. Therefore, the power consumed by each $(feature,channel)$ of the C-Elements ($P_{C-Element}$) can be estimated as
\begin{equation}
    P_{C-Element} =  1000\ \text{Hz} \cdot 20\ \text{fJ} = 20\ \text{pW}.
\end{equation}

\subsection{Neurostimulation}

According to \cite{Sun2014, NeuroPace}, the typical responsive stimulation to seizures is a series of pulses with a frequency of $200 Hz$, individual pulse-widths of $160 \mu s$, $6 mA$ amplitude, and total active duration of 100 ms. For a 1200 ohm lead resistance, each stimulation activity consumes:
\begin{equation}
    E_{stim, avg} = I_{stim}^2 \cdot R_{lead} \cdot T \cdot D = 138 \mu J,
\end{equation}

where $E_{stim, avg}$ is the energy required to generate a series of pulses for one stimulation activity with average strength; $I_{stim}$ is the stimulating current amplitude; $R_{lead}$ is the lead resistance; $T$ is the duration of one stimulation activity; and $D = 160 \mu s \cdot 200 Hz = 3.2\%$ is the duty cycle.

When tuned for the patients who need more stimulations, with $I_stim = 12 mA$, and a $200 \mu s$ pulse width, we get:
\begin{equation}
    E_{stim, high} = 691 \mu J,
\end{equation}

\subsection{Total Power Estimation}

The total power consumption per $(feature,channel)$ for the detection system is:

\begin{multline}
    P_{total} = P_{sen} + P_{pre-amp} + P_{FE-avg} + P_{LUT}\\ + P_{TRNG} + P_{C-Element} = 6.189\  \mu \text{W}
\end{multline}

Here, $P_{sen}$, $P_{pre-amp}$, $P_{FE-avg}$, $P_{LUT}$, $P_{TRNG}$, and $P_{C-Element}$ represent the power consumption on sensing, pre-amplification, (average) feature extraction, look-up table, true random number generator, and C-element blocks respectively.  Ultimately, our detection system uses about $6.189 \mu W$ of power per $(feature,channel)$ for detection.  Preliminary simulation results indicate that detection accuracies will continue to improve with an increasing number of $(feature,channel)$ pairs. Assuming that an average of four $(feature,channel)$ pairs are used per patient, the detection circuitry consumes roughly $25\mu W$.  For stimulation, average seizure-activity patients will use $138 \mu J$ per stimulation 570 times a day giving a power consumption of $0.9 \mu W$.  Similarly, high seizure-activity patients use $691 \mu J$ per stimulation 1330 times a day giving a power consumption of $11 \mu W$ Assuming a battery capacity of $3.3 Wh$ \cite{NeuroPace}, the battery for low-activity patients will last over 14 years, and the battery for high activity patients will last over 10 years, both of which are significant improvements from the numbers reported by NeuroPace.

\section{Conclusions and Future Work}

We have demonstrated preliminary results for the implementations of a very-low-power seizure detection engine which will simultaneously increase detection accuracies, decrease false positives and over-stimulation, and boost battery life by up to 50\% as compared with state-of-the-art responsive neurostimulation systems.  Each of these improvements will increase a patient's quality of life.

Future work will be dedicated toward refining the preliminary results presented herein including: performing full SPICE simulations of the system tuned to specific patients, using less-noisy inter-cranial EEG data, and aiming for a fabricated system.  Ultimately, we believe we have demonstrated the plausibility of our system and shown its potential to improve the quality of life for patients suffering from epilepsy.


\bibliographystyle{IEEEtran}
\bibliography{main}

\begin{thebibliography}{10}
\providecommand{\url}[1]{#1}
\csname url@samestyle\endcsname
\providecommand{\newblock}{\relax}
\providecommand{\bibinfo}[2]{#2}
\providecommand{\BIBentrySTDinterwordspacing}{\spaceskip=0pt\relax}
\providecommand{\BIBentryALTinterwordstretchfactor}{4}
\providecommand{\BIBentryALTinterwordspacing}{\spaceskip=\fontdimen2\font plus
\BIBentryALTinterwordstretchfactor\fontdimen3\font minus
  \fontdimen4\font\relax}
\providecommand{\BIBforeignlanguage}[2]{{%
\expandafter\ifx\csname l@#1\endcsname\relax
\typeout{** WARNING: IEEEtran.bst: No hyphenation pattern has been}%
\typeout{** loaded for the language `#1'. Using the pattern for}%
\typeout{** the default language instead.}%
\else
\language=\csname l@#1\endcsname
\fi
#2}}
\providecommand{\BIBdecl}{\relax}
\BIBdecl

\bibitem{WHO}
\BIBentryALTinterwordspacing
W.~H. Organisation. (2019, June) Epilepsy. [Online]. Available:
  \url{https://www.who.int/news-room/fact-sheets/detail/epilepsy}
\BIBentrySTDinterwordspacing

\bibitem{drug_resistant_epilepsy_2018}
aul Boon \emph{et~al.}, ``\BIBforeignlanguage{English}{Neurostimulation for
  drug-resistant epilepsy: A systematic review of clinical evidence for
  efficacy, safety, contraindications and predictors for response},''
  \emph{\BIBforeignlanguage{English}{Current Opinion in Neurology}}, vol.~31,
  no.~2, pp. 198--210, apr 2018.

\bibitem{Assi2017}
\BIBentryALTinterwordspacing
E.~{Bou Assi} \emph{et~al.}, ``Towards accurate prediction of epileptic
  seizures: A review,'' \emph{Biomedical Signal Processing and Control},
  vol.~34, pp. 144--157, 2017. [Online]. Available:
  \url{https://www.sciencedirect.com/science/article/pii/S1746809417300277}
\BIBentrySTDinterwordspacing

\bibitem{Neuromodulation_treatment_2020}
\BIBentryALTinterwordspacing
G.~J. Davis~P, ``Neuromodulation for the treatment of epilepsy: A review of
  current approaches and future directions.'' \emph{Clinical therapeutics},
  June 2020. [Online]. Available:
  \url{https://doi.org/10.1016/j.clinthera.2020.05.017}
\BIBentrySTDinterwordspacing

\bibitem{invasive_neurostimulation_epilepsy_2020}
S.~R. Alomar~SA, ``Different modalities of invasive neurostimulation for
  epilepsy.'' \emph{Neurological Sciences.}, August 2020.

\bibitem{Friedman_C-Element_BI-2016}
J.~S. Friedman \emph{et~al.}, ``Bayesian inference with muller c-elements,''
  \emph{IEEE Transactions on Circuits and Systems I: Regular Papers}, vol.~63,
  no.~6, 2016.

\bibitem{NeuroPace}
\BIBentryALTinterwordspacing
NeuroPace. (2020, 6) {RNS} system physician manual. [Online]. Available:
  \url{https://www.neuropace.com/wp-content/uploads/2021/02/neuropace-rns-system-manual-320.pdf}
\BIBentrySTDinterwordspacing

\bibitem{AFullyInteg8ChCL-Chen-2014}
W.-M. Chen \emph{et~al.}, ``A fully integrated 8-channel closed-loop
  neural-prosthetic cmos soc for real-time epileptic seizure control,''
  \emph{IEEE Journal of Solid-State Circuits}, vol.~49, no.~1, pp. 232--247,
  2014.

\bibitem{APersonalEAIEEG-Pinto-2021}
\BIBentryALTinterwordspacing
M.~F. Pinto \emph{et~al.}, ``A personalized and evolutionary algorithm for
  interpretable eeg epilepsy seizure prediction,'' \emph{Scientific Reports},
  vol.~11, no.~1, p. 3415, Feb 2021. [Online]. Available:
  \url{https://doi.org/10.1038/s41598-021-82828-7}
\BIBentrySTDinterwordspacing

\bibitem{An8ChannelSEEGASWPS-Yoo-2013}
J.~Yoo \emph{et~al.}, ``An 8-channel scalable eeg acquisition soc with
  patient-specific seizure classification and recording processor,'' \emph{IEEE
  Journal of Solid-State Circuits}, vol.~48, no.~1, pp. 214--228, 2013.

\bibitem{A16ChannelPSSOaT-Altaf-2015}
M.~A. Bin~Altaf \emph{et~al.}, ``A 16-channel patient-specific seizure onset
  and termination detection soc with impedance-adaptive transcranial electrical
  stimulator,'' \emph{IEEE Journal of Solid-State Circuits}, vol.~50, no.~11,
  pp. 2728--2740, 2015.

\bibitem{stochastic-computing}
B.~R. Gaines, \emph{Stochastic Computing Systems}.\hskip 1em plus 0.5em minus
  0.4em\relax Boston, MA: Springer US, 1969, pp. 37--172.

\bibitem{HOE20194}
D.~H. Hoe, ``Bayesian inference using stochastic logic: A study of buffering
  schemes for mitigating autocorrelation,'' \emph{International Journal of
  Approximate Reasoning}, vol. 112, pp. 4--21, 2019.

\bibitem{Zhou2008}
\BIBentryALTinterwordspacing
S.~hua Zhou \emph{et~al.}, ``An ultra-low power cmos random number generator,''
  \emph{Solid-State Electronics}, vol.~52, no.~2, pp. 233--238, 2008. [Online].
  Available:
  \url{https://www.sciencedirect.com/science/article/pii/S003811010700295X}
\BIBentrySTDinterwordspacing

\bibitem{Yang2015}
K.~Yang \emph{et~al.}, ``A robust -40 to 120°c all-digital true random number
  generator in 40nm cmos,'' in \emph{2015 Symposium on VLSI Circuits (VLSI
  Circuits)}, 2015, pp. C248--C249.

\bibitem{Mathew2016}
S.~K. Mathew \emph{et~al.}, ``$\mu $ rng: A 300–950 mv, 323 gbps/w
  all-digital full-entropy true random number generator in 14 nm finfet cmos,''
  \emph{IEEE Journal of Solid-State Circuits}, vol.~51, no.~7, pp. 1695--1704,
  2016.

\bibitem{Vodenicarevic2017}
\BIBentryALTinterwordspacing
D.~Vodenicarevic \emph{et~al.}, ``Low-energy truly random number generation
  with superparamagnetic tunnel junctions for unconventional computing,''
  \emph{Phys. Rev. Applied}, vol.~8, p. 054045, Nov 2017. [Online]. Available:
  \url{https://link.aps.org/doi/10.1103/PhysRevApplied.8.054045}
\BIBentrySTDinterwordspacing

\bibitem{P-Bit_generators-RFaria-2017}
\BIBentryALTinterwordspacing
K.~Y. Camsari \emph{et~al.}, ``Stochastic $p$-bits for invertible logic,''
  \emph{Phys. Rev. X}, vol.~7, p. 031014, Jul 2017. [Online]. Available:
  \url{https://link.aps.org/doi/10.1103/PhysRevX.7.031014}
\BIBentrySTDinterwordspacing

\bibitem{Camsari2017-Stochastic-P-Bits-MTJ}
------, ``Implementing p-bits with embedded mtj,'' \emph{IEEE Electron Device
  Letters}, vol.~38, no.~12, pp. 1767--1770, 2017.

\bibitem{TUH-Database}
V.~Shah \emph{et~al.}, ``The temple university hospital seizure detection
  corpus,'' \emph{Frontiers in Neuroinformatics}, vol.~12, p.~83, 2018.

\bibitem{Shoaran2018_EnEffClass_5FoldCV}
M.~Shoaran \emph{et~al.}, ``{Energy-efficient classification for
  resource-constrained biomedical applications},'' \emph{IEEE Journal on
  Emerging and Selected Topics in Circuits and Systems}, vol.~8, no.~4, pp.
  693--707, dec 2018.

\bibitem{TRUONG2018-ConvNN}
\BIBentryALTinterwordspacing
N.~D. Truong \emph{et~al.}, ``Convolutional neural networks for seizure
  prediction using intracranial and scalp electroencephalogram,'' \emph{Neural
  Networks}, vol. 105, pp. 104--111, 2018. [Online]. Available:
  \url{https://www.sciencedirect.com/science/article/pii/S0893608018301485}
\BIBentrySTDinterwordspacing

\bibitem{Daoud2019-EESeizPredDeepLearn}
H.~Daoud \emph{et~al.}, ``Efficient epileptic seizure prediction based on deep
  learning,'' \emph{IEEE Transactions on Biomedical Circuits and Systems},
  vol.~13, no.~5, pp. 804--813, 2019.

\bibitem{Yang2020-SeizRev}
J.~Yang \emph{et~al.}, ``From seizure detection to smart and fully embedded
  seizure prediction engine: A review,'' \emph{IEEE Transactions on Biomedical
  Circuits and Systems}, vol.~14, pp. 1008--1023, 08 2020.

\bibitem{MullerAmplifier2012}
R.~Muller \emph{et~al.}, ``A 0.013 ${\hbox {mm}}^{2}$, 5 $\mu\hbox{W}$ ,
  dc-coupled neural signal acquisition ic with 0.5 v supply,'' \emph{IEEE
  Journal of Solid-State Circuits}, vol.~47, no.~1, pp. 232--243, 2012.

\bibitem{Xu2008}
X.~Xu \emph{et~al.}, ``A 1-v 450-nw fully integrated biomedical sensor
  interface system,'' in \emph{2008 IEEE Symposium on VLSI Circuits}, 2008, pp.
  78--79.

\bibitem{Zou2010}
X.~Zou \emph{et~al.}, ``A 1v 22µw 32-channel implantable eeg recording ic,''
  in \emph{2010 IEEE International Solid-State Circuits Conference - (ISSCC)},
  2010, pp. 126--127.

\bibitem{Denison2007}
T.~Denison \emph{et~al.}, ``A 2.2/spl mu/w 94nv//spl radic/hz,
  chopper-stabilized instrumentation amplifier for eeg detection in chronic
  implants,'' in \emph{2007 IEEE International Solid-State Circuits Conference.
  Digest of Technical Papers}, 2007, pp. 162--594.

\bibitem{Sun2014}
\BIBentryALTinterwordspacing
F.~T. Sun \emph{et~al.}, ``The rns system: responsive cortical stimulation for
  the treatment of refractory partial epilepsy,'' \emph{Expert Review of
  Medical Devices}, vol.~11, no.~6, pp. 563--572, 2014, pMID: 25141960.
  [Online]. Available: \url{https://doi.org/10.1586/17434440.2014.947274}
\BIBentrySTDinterwordspacing

\end{thebibliography}

\end{document}